%% file: apssamp.tex
\documentclass[
 reprint,
superscriptaddress,
 amsmath,amssymb,
 aps
]{revtex4-2}

\usepackage{graphicx}
\usepackage{dcolumn}
\usepackage{bm}
\usepackage{hyperref}
\usepackage{mathtools}
\usepackage{braket}
\usepackage{xcolor}  
\usepackage{csquotes}
\usepackage{siunitx}

\begin{document}

\preprint{APS/123-QED}

\title{Sparse Weyl Propagation}

\author{Yanick S. Kind}
\affiliation{Condensed Matter Theory, TU Dortmund University, Otto-Hahn-Straße 4, 44227 Dortmund, Germany}

\author{Benedikt Fauseweh}
 \email{benedikt.fauseweh@tu-dortmund.de}
\affiliation{Condensed Matter Theory, TU Dortmund University, Otto-Hahn-Straße 4, 44227 Dortmund, Germany}

\date{\today}
\begin{abstract}
We investigate the generalization of sparse Pauli propagation to qudits with $d$ levels by propagating observables backwards in the basis of Weyl strings.
We show that a Weyl string is mapped to at most $d$ Weyl strings under gates generated by Weyl strings and provide a closed form update rule for arbitrary local dimensions $d$. The truncation error has an upper bound in terms of the operator stabilizer R\'enyi entropy.
Performing numerical simulations for the kicked two-dimensional $Z_3$ clock model and the field quenched toric-code, we benchmark sparse Weyl propagation against state-of-the-art tensor network simulations and against exact diagonalization for small systems. 
We further show results for systems out of reach for exact diagonalization.
We find that this approach can yield results orders of magnitude more precise than tensor network simulations, especially in strongly entangled regimes.
\end{abstract}

\maketitle

\input{content/introduction}
\input{content/results}

\input{content/discussion_outlook}\\ 

\textit{Acknowledgments}---This work was funded by the Deutsche Forschungsgemeinschaft (DFG, German Research Foundation)-Project No. FA 1884/5-1. The authors formulated the research questions and developed the research approach without AI assistance.
Software and quantum circuits were implemented using Claude Opus 5.5.
OpenAI ChatGPT 6.1 Astra and Claude Opus 5.5 were additionally used to check generated code for bugs. 
The authors designed the manuscript and used Claude Opus 5.5 to refine wording and sentence structure and for literature research. \\

\textit{Data availability}---The data that support the findings of
this article are openly available~\cite{kind_2026_24585838}.

\bibliography{apssamp}
\clearpage          
\onecolumngrid
\appendix
\input{content/appendix}

\end{document}

%% file: content/introduction.tex
\textit{Introduction}--
Simulating the dynamics of quantum many-body systems on a classical computer is central to understanding
non-equilibrium phenomena and benchmarking quantum processors~\cite{kim2023, PRXQuantum.5.010308, doi:10.1126/sciadv.adk4321, Fauseweh2024, fhc5-8sm6, barron2026observableestimationabsenceclassical}. 
However, the exponential scaling of the Hilbert space dimension with system size remains prohibitive for exact computations so that one has to resort to some form of compression.

Strong candidates for simulating large quantum systems are tensor network methods, which compress the state into a network of local tensors.
Their bond dimension $\chi$ bounds the entanglement entropy that can be encoded by the network.
In two dimensions, these methods require contracting loopy networks. Exact contraction is exponentially costly, but can be approximated by belief propagation (BP)~\cite{SciPostPhys.15.6.222, PhysRevResearch.3.023073}.

Sparse Pauli propagation (SPP)~\cite{10.1063/5.0269149,6vd7-l9bn, PhysRevA.99.062337, PRXQuantum.6.020302, lh6x-7rc3}, on the other hand, resorts to compressing the observable. 
Here, instead of forward evolving the state by a unitary circuit, the observable is propagated backwards through the circuit in the Heisenberg picture.
Its accuracy is not limited by entanglement, but by the operator magic of the evolved observable~\cite{shao2026,p7xt-s9nz}.
For a circuit consisting only of Clifford gates, the number of Pauli operators representing an Observable does not increase, and no truncation is needed. 
However, recent progress~\cite{PRXQuantum.6.020302,doi:10.1126/sciadv.adk4321} showed that SPP can accurately capture the dynamics of quantum systems even if the circuit does not purely consist of Clifford gates and a truncation is required.

SPP relies on the decomposition of operators in the Pauli operator basis so that the method is limited to qubits. 
However, many platforms natively operate on $d$-level systems, i.e., qudits~\cite{10.3389/fphy.2020.589504, ringbauer2022, hrmo2023, PhysRevX.11.021010, PhysRevLett.126.210504, goss2022, chi2022, lib2026qutritentanglementjointmultiparameter}. 

In this paper, we introduce sparse Weyl propagation (SWP), a generalization of SPP to qudits. 
We derive an update rule for gates generated by arbitrary Weyl strings. 
We benchmark SWP against projected entangled pair states (PEPS) with BP in a kicked $\mathbb{Z}_3$ clock model discrete time crystal~\cite{PhysRevB.99.104303} and a $\mathbb{Z}_3$ toric code~\cite{KITAEV20032,Bullock_2007} for qutrits.
We show results for system sizes far beyond the reach of exact diagonalization (ED), while for small systems, both methods are validated against ED.

Weyl methods are already used to describe quantum processes to  learn error models~\cite{PhysRevLett.126.210502} but not to simulate the dynamics of quantum systems. So far they have also not been compared to tensor network methods.

For qudits with $d$ levels $0,1 \ldots, d-1$ the clock operator $Z$ and shift operator $X$ are defined as $X\ket{j} = \ket{(j+1)\,\bmod\,d}$ and $Z \ket{j} = \omega^j \ket{j}$ with $\omega = \exp(i2\pi/d)$. 
The operators commute up to the phase $\omega$, i.e., $ZX = \omega XZ$.
Furthermore they fulfill the identities $X^d = Z^d = \openone$,
${X^\dagger = X^{-1} = X^{d-1}}$ and ${Z^\dagger = Z^{-1} = Z^{d-1}}$.
The products $W_{(x,z)} = X^x Z^z, \; (x,z)\in \mathbb{Z}_d\times\mathbb{Z}_d$ acting on a basis state as $W_{(x,z)}\ket{j}=\omega^{zj}\ket{(j+x)\bmod d}$ are the single qudit  \textit{Weyl operators}, of which $d^2$ exist. 
Note that for $d=2$ there are $\openone, X, Z, XZ$, i.e. the Pauli-matrices are recovered up to a phase since $XZ = -iY$ ($\omega = -1$ for qubits).
Multiplying two Weyl operators yields $W_{(x,z)} W_{(x',z')} = \omega^{x'z}W_{(x+x', z+z')}$.
An important fact is that nontrivial Weyl operators are traceless 
$\text{Tr}[W_{(x,z)}] = \sum_{j=0}^{d-1} \braket{j| X^xZ^z |j}$ = $ \sum_{j=0}^{d-1} \omega^{zj}\braket{j|(j+x)\, \bmod \, d} = \sum_{j=0}^{d-1} \omega^{zj} \delta_{j,(j+x)\, \bmod \, d}=\delta_{x,0} \sum_{j=0}^{d-1} \omega^{zj} = d \delta_{x,0} \delta_{z,0}$,
since $\sum_{j=0}^{d-1} \omega^{zj} =0$ for $z \neq 0$.
Using the identity for the trace together with the rule for multiplying two Weyl operators we obtain the orthogonality relation $\text{Tr}[W^{\dagger}_{(x,z)} W_{(x',z')}] = d \delta_{x,x'} \delta_{z,z'}$.
Since there are $d^2$ orthogonal Weyl operators in a $d^2$-dimensional space, they form an \textit{orthogonal operator basis}.
We can therefore expand every single qudit operator in the Weyl operator basis  
\begin{equation}
    O = \sum_{(x,z)} c_{(x,z)}(O) W_{(x,z)}, \quad c_{(x,z)}(O) = d^{-1} \, \text{Tr}[W_{(x,z)}^{\dagger} O].
\end{equation}
Note that for qubits the phase convention $Y=iXZ$ yields a Hermitian operator basis ${\openone, X,Z,Y}$ such that all coefficients are real for an Observable $O$. For $d>2$ the operator basis is not Hermitian and the coefficients $c_{(x,z)}(O)$ are complex in general.
For $n$ qudits the basis states are the tensor products of the single qudit basis states $\ket{j_1}\otimes \ldots \otimes \ket{j_n} \equiv \ket{j_1, \ldots, j_n} =  \ket{\mathbf{j}}$ with $\mathbf{j} \in \mathbb{Z}_d^n$,  and the corresponding \textit{Weyl strings} are labeled by the phase space points $a = ((x_1, \ldots, x_n), (z_1, \ldots , z_n)) \equiv (\mathbf{x}, \mathbf{z}) \in \mathbb{Z}_d^n \times \mathbb{Z}_d^n$
\begin{equation}
  W_{(\mathbf{x}, \mathbf{z})} \coloneqq X^{x_1} Z^{z_1} \otimes \cdots \otimes X^{x_n} Z^{z_n} = \bigotimes_{i=1}^n X^{x_i}Z^{z_i}.
  \label{eq:W}   
\end{equation}
The identity $\text{Tr}[W^{\dagger}_a W_b] = d^n \delta_{a,b}$ factorizes into single-qudit traces.
Together with the fact that there are $d^{2n}$ Weyl strings we can conclude that the Weyl strings, again, form an orthogonal operator basis for $n$ qudit operators
so that we can expand any $n$ qudit operator $O$ in terms of Weyl strings 
\begin{equation}
    O = \sum_a c_a (O) W_a, \quad c_a(O) = d^{-n} \text{Tr}[W_a^{\dagger} O].
\end{equation}
Multiplying two Weyl strings yields 
\begin{equation}
    W_{(\mathbf{x},\mathbf{z})} W_{(\mathbf{x}',\mathbf{z}')} = \omega^{\mathbf{x}'\cdot \mathbf{z}} W_{(\mathbf{x}+\mathbf{x}',\mathbf{z}+\mathbf{z}')},
    \label{eqn:mul}
\end{equation}
which is particularly important for propagating observables backwards through the circuit. 
Note that the commutation relation
$W_{(\mathbf{x},\mathbf{z})} W_{(\mathbf{x}',\mathbf{z}')} = \omega^{\mathbf{x}'\cdot\mathbf{z}-\mathbf{x}\cdot\mathbf{z}'}W_{(\mathbf{x}',\mathbf{z}')} W_{(\mathbf{x},\mathbf{z})}$
follows.

The goal of SWP is to compute expectation values of the form 
\begin{equation}
    E = \braket{\psi_0 |U_1^\dagger\dots U_L^\dagger O U_L \dots U_1|\psi_0},
\end{equation}
where $\ket{\psi_0}$ is an initial state, $O$ an observable, and each $U_i$
is a gate generated by a Weyl string,
\begin{equation}
  U_{(\mathbf{x},\mathbf{z})}
  = \exp\!\big[-i\theta\big(e^{i\vartheta}W_{(\mathbf{x},\mathbf{z})}
      + e^{-i\vartheta}W^\dagger_{(\mathbf{x},\mathbf{z})}\big)\big],
  \qquad \theta,\vartheta\in\mathbb{R}.
  \label{eq:gate}
\end{equation}
Since Weyl strings are not Hermitian for $d>2$, a Hermitian linear combination in
the exponent is required for $U$ to be unitary.
Now instead of forward evolving the state $\ket{\psi_i} = U_i \dots U_1 \ket{\psi_0}$ in the Schrödinger picture, SWP propagates the observable backwards through the circuit in the Heisenberg picture 
$O_i =  U_i^\dagger \dots U_L^\dagger O U_L \dots U_i$.
Since $W_{(\mathbf{x},\mathbf{z})}^d \propto \openone$, any gate of the form~\eqref{eq:gate} is a polynomial in $W_{(\mathbf{x},\mathbf{z})}$,
$U_{(\mathbf{x},\mathbf{z})} = \sum_{m=0}^{d-1} f_m W_{(\mathbf{x},\mathbf{z})}^m$. 
Thus, conjugating a single Weyl string $V_{(\mathbf{x}',\mathbf{z}')}$ with a gate $U_{(\mathbf{x},\mathbf{z})}$ yields 
\begin{align}
  U_{(\mathbf{x},\mathbf{z})}^\dagger V_{(\mathbf{x}',\mathbf{z}')} U_{(\mathbf{x},\mathbf{z})} &= \sum_{m=0}^{d-1} \beta_m\, V_{(\mathbf{x}',\mathbf{z}')} W_{(\mathbf{x}, \mathbf{z})}^{m},  \label{eqn:gate_application} \\
  \beta_m = \sum_{p=0}^{d-1} f_p^*\,  & f_{(m+p) \bmod d} \; 
  \omega^{-p\sigma}\; \gamma_{(\mathbf x,\mathbf z)}^{\lfloor (m+p)/d \rfloor} \nonumber
\end{align}
with $\sigma = \mathbf{x}' \cdot \mathbf{z}- \mathbf{x}\cdot \mathbf{z}'$ and $f_p$ derived in Appendix~\ref{app:update} (see Eq.~\eqref{eqn:coeffs}).
From Eq.~\eqref{eqn:mul}, $W_{(\mathbf{x}, \mathbf{z})}^d = \gamma_{(\mathbf{x}, \mathbf{z})} \openone$ with
$\gamma_{(\mathbf{x}, \mathbf{z})} = \omega^{\frac{d(d-1)}{2}\mathbf{x}\cdot\mathbf{z}}
= (-1)^{(d-1)\mathbf{x}\cdot\mathbf{z}}$,
i.e.\ $\gamma_{(\mathbf{x}, \mathbf{z})} = 1$ for odd $d$ and $\gamma_{(\mathbf{x}, \mathbf{z})} = (-1)^{\mathbf{x}\cdot\mathbf{z}}$ for even $d$.
Eq.~\eqref{eqn:gate_application} shows that a single Weyl string branches into at most $d$ Weyl strings.
In case $V_{(\mathbf{x}',\mathbf{z}')}$ and $W_{(\mathbf{x},\mathbf{z})}$ commute, $\beta_m = \delta_{m,0}$ and the Weyl string is unchanged.

A general observable is a superposition of Weyl strings $O=\sum_a c_a(O)W_a$. 
By linearity, each gate is applied to every Weyl string via Eq.~\eqref{eqn:gate_application} and the coefficients of identical Weyl strings are added. 
Without truncation the number of Weyl strings grows exponentially. 
In our implementation, we keep the Weyl strings with the $K$ largest coefficients by magnitude, i.e., Top-$K$ truncation. 
Note that the truncation affects the Hilbert-Schmidt norm so that the norm of the discarded part serves as an error indicator.
The parameter $K$ controls the trade-off between computational cost and accuracy.
For a product state $\ket{\psi_0}$, the expectation value of a single Weyl string factorizes into single qudit matrix elements. 
If $\ket{\psi_0}$ is a stabilizer state, such as the ground states of the toric code,  $\braket{\psi_0 | W_a | \psi_0}$ is only non-zero if $W_a$ belongs to the stabilizer group up to a phase.
Per gate, the runtime and memory costs are $\mathcal{O}(dK)$.

The error of the Top-$K$ truncation can be rigorously bounded in terms of the operator stabilizer R\'enyi entropy (OSE) of the observable~\cite{p7xt-s9nz,shao2026}.
With the normalized Hilbert-Schmidt norm $\|O\|^2 = d^{-n}\,\text{Tr}[O^\dagger O] = \sum_a |c_a(O)|^2$, the weights $p_a = |c_a(O)|^2/\|O\|^2$ form a probability distribution over the Weyl strings, and the OSE of order $\alpha$ is its R\'enyi entropy
\begin{equation}
  \mathcal{S}_\alpha(O) = \frac{1}{1-\alpha}\ln\sum_a p_a^\alpha .
  \label{eqn:ose}
\end{equation}
It vanishes for a single Weyl string and is invariant under Clifford gates, which map Weyl strings onto Weyl strings up to phases.
Generalizing the argument of Ref.~\cite{shao2026} to qudits, we sort the weights, $p_{(1)}\geq p_{(2)}\geq\dots$, and use $K p_{(K)}^\alpha\leq\sum_a p_a^\alpha$.
For $0<\alpha<1$, the relative weight discarded by a Top-$K$ truncation $\mathcal{T}_K$ is then bounded by
\begin{align}
  \frac{\|O-\mathcal{T}_K(O)\|^2}{\|O\|^2}
  &= \sum_{k>K} p_{(k)} \leq p_{(K)}^{1-\alpha}\sum_{k>K} p_{(k)}^{\alpha} \nonumber\\
  &\leq e^{\frac{1-\alpha}{\alpha}\left[\mathcal{S}_\alpha(O)-\ln K\right]} .
  \label{eqn:tail}
\end{align}
Since gates preserve the norm and truncations do not increase it, the errors of the $L$ truncations add up at most linearly,
\begin{equation}
  \|O_1-\tilde{O}_1\|
  \leq \|O\|\sum_{i=1}^{L} e^{\frac{1-\alpha}{2\alpha}\left[\mathcal{S}_\alpha(U_i^\dagger\tilde{O}_{i+1}U_i)-\ln K\right]} ,
  \label{eqn:apriori}
\end{equation}
where $\tilde{O}_i=\mathcal{T}_K(U_i^\dagger\tilde{O}_{i+1}U_i)$ with $\tilde{O}_{L+1}=O$ is the truncated observable.\\

%% file: content/results.tex
\begin{figure*}[t]
    \includegraphics{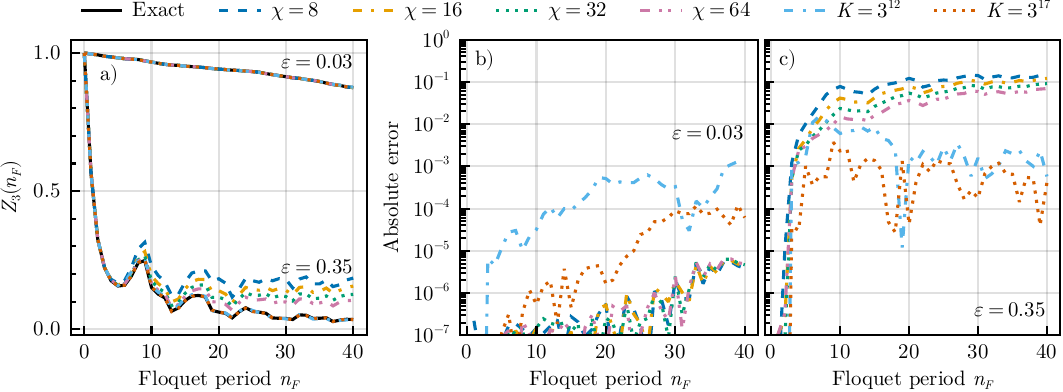}
    \caption{\label{fig:clock4x4}Simulation results for the clock model with $L_x=L_y=4$, period $T=1$, chirality $\phi = \pi / 6$, random couplings $J_{s,s'}\in [1/2, 3/2]$ and random fields $h_s \in [0,0.3]$ with different bond dimension $\chi$ for the tensor network simulations and different top-$K$ truncations for the SWP. In a) we report the $Z_3$ order parameter as defined in Eq.~\eqref{eqn:z3} and the absolute error with respect to the exact calculations for b) the DTC phase and for c) the melted phase.}
    \label{fig:clock_ed}
\end{figure*}
\begin{figure*}[t]
    \includegraphics{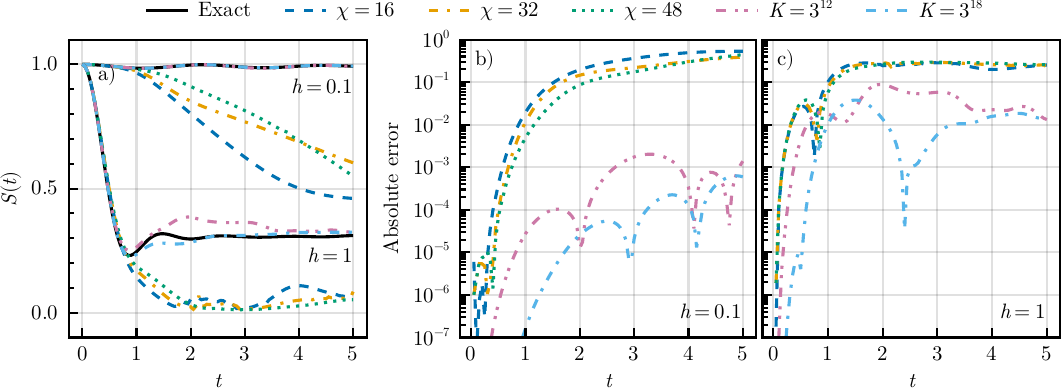}
    \caption{\label{fig:toric_3x3}Simulation results for the toric code with $L_x=L_y=3$, $h_{x,e}, h_{z,e}\in [0.8h,1.2h]$ and time step $\delta t = 0.05$ with different bond dimension $\chi$ for the tensor network simulations and different top-$K$ truncations for the SWP. In a) we report the average star expectation value $S(t)$ and the absolute error with respect to the exact calculations for b) the weak quench phase and c) for the strong quench.}
    \label{fig:toric_ed}
\end{figure*}
\textit{Results}--Our first model is a two-dimensional $\mathbb{Z}_3$ kicked clock model on an open $L_x \times L_y$ square lattice, i.e. 
$n=L_x L_y$ qutrits.
The Floquet unitary of one period is given by
\begin{align}
  U_F &= e^{-iTH} K_\varepsilon  \label{eq:UF} \\
  K_\varepsilon &= \prod_s X_s \, e^{i\varepsilon \Theta_s} ,
  \quad 
  \Theta_s = \frac{2i}{\sqrt{3}} \bigl( X_s^\dagger - X_s \bigr) ,
  \label{eq:Theta} \\
  H &= \bigl( \sum_{\langle s  s' \rangle} J_{ss'} e^{i\phi} Z_s^\dagger Z_{s'} 
  + \sum_s h_s  Z_s , \bigr) + \text{h.c.}
  \label{eq:H}
\end{align}
with chirality $\phi = \pi/6$, period $T = 1$, and disordered couplings
$J_{ss'} \in [1/2,\, 3/2]$ and fields $h_s \in [0,\, 0.3]$,
where $\langle s  s' \rangle$ denotes the oriented nearest-neighbor bonds.

In the time-crystalline phase \cite{PhysRevB.99.104303,Camacho_2026}, the magnetization rotates by $e^{2\pi i/3}$
each Floquet period $n_F$. 
We therefore track the $Z_3$ order parameter
\begin{align}
  Z_3(n_F) &= \operatorname{Re}\bigl[ e^{-2\pi i n_F/3}\, m(n_F) \bigr] ,
  \label{eqn:z3}\\
  m(n_F) &= \frac{1}{n} \sum_s \langle \psi_0 | U_F^{\dagger n_F} Z_s U_F^{n_F} | \psi_0 \rangle ,
  \nonumber
\end{align}
starting from $|\psi_0\rangle = |0\rangle^{\otimes n}$. 
Perfect period-3 order (${\varepsilon=0}$) gives $Z_3 = 1$, while $Z_3 \to 0$ when the order melts.

The second model is the toric code on a periodic ${L_x\times L_y}$ square lattice, where the qutrits live on the edges. 
Thus there are $n = 2L_x L_y$ qutrits.
We label the vertices $\nu (x,y)$ by the coordinates $x=0,\ldots,L_x -1$, $y=0,\ldots,L_y -1$, the horizontal edges by $e_x(x,y)$ and the vertical edges by $e_y(x,y)$.
The star operators are defined by $A_{x,y}=X_{e_x (x,y)} X_{e_y (x,y)} X^{\dagger}_{e_x (x-1,y)} X^{\dagger}_{e_y (x,y-1)}$, while the plaquette operators are defined by $B_{x,y} = Z_{e_x (x,y)} Z_{e_y (x+1,y)} Z^{\dagger}_{e_x (x,y+1)} Z^{\dagger}_{e_y (x,y)}$.
We prepare a ground state of the toric code Hamiltonian $
H_0 = -\frac{1}{2} \sum_v \bigl( A_v + A_v^\dagger \bigr) - \frac{1}{2} \sum_p \bigl( B_p + B_p^\dagger \bigr) \label{eqn:H0}$.
The ground state manifold of $H_0$ is nine-fold degenerate. 
We choose one ground state $\ket{\psi_0}$ by fixing the two mutually commuting, noncontractible logical $Z$-loop operators
to eigenvalue $+1$, $\bar{Z}_x \ket{\psi_0} = |\psi_0\rangle$ and $\bar{Z}_y \ket{\psi_0} = \ket{\psi_0}$.
The system is quenched at $t=0$ by switching on fields on all edges so that the generator of the time evolution here is
\begin{align}
  H = & \underbrace{-\frac{1}{2} \sum_{\nu} \bigl( A_v + A_v^\dagger \bigr)
         - \sum_e \frac{h_{x,e}}{2} \bigl( X_e + X_e^\dagger \bigr) }_{H_\text{A}}
  \nonumber \\
        & \underbrace{- \frac{1}{2} \sum_p \bigl( B_p + B_p^\dagger \bigr)
         - \sum_e \frac{h_{z,e}}{2} \bigl( Z_e + Z_e^\dagger \bigr)}_{H_\text{B}},
  \label{eq:Hf}
\end{align}
where $e$ labels the edges and $p$ the plaquettes.
We incorporate disorder into the fields by choosing $h_{x,e} = h (1 + \delta_e)$ and $h_{z,e} = h (1 + \delta'_e)$
with $\delta_e, \delta'_e$ drawn uniformly from $[-0.2, 0.2]$.
The time evolution is implemented by using a second order Trotter-Suzuki decomposition $\text{exp}(-iH\delta t) = \text{exp}(-iH_\text{A}\delta t/2)\text{exp}(-iH_\text{B}\delta t)\text{exp}(-iH_\text{A}\delta t/2) + \mathcal{O}(\delta t^3)$.
All terms within  $H_\text{A}$ and $H_\text{B}$ mutually commute so that no further approximation is needed for the time evolution.
We track the average star expectation value $S(t)=2n^{-1} \sum_v \operatorname{Re} \braket{\psi_0|U^\dagger(t) A_v U(t) | \psi_0}$.

For details on the implementation we refer the reader to Appendix \ref{app:simulation}.

\begin{figure}
    \includegraphics{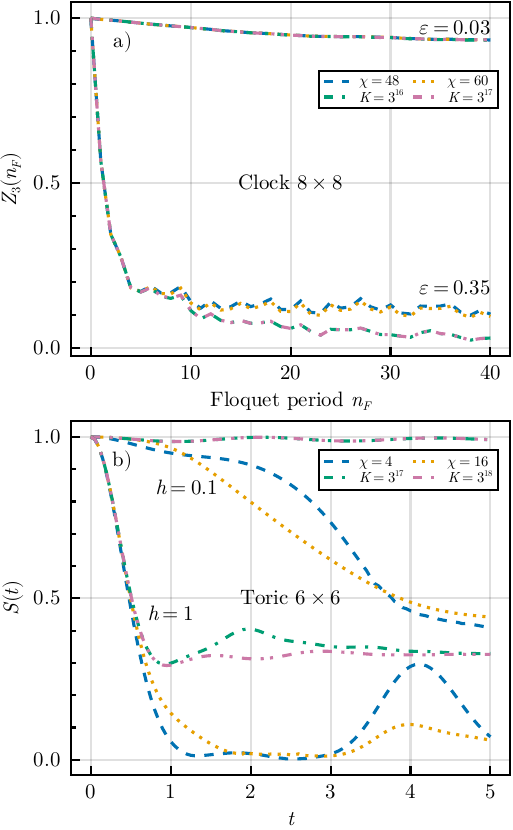}
    \caption{Simulation results for both models for system sizes beyond the reach of ED with different bond dimension $\chi$ for the tensor network simulations and different top-$K$ truncations for the SWP. In a) we report the $Z_3$ order parameter as defined in Eq.~\eqref{eqn:z3} for the clock model with $L_x=L_y=8$, period $T=1$, chirality $\phi = \pi / 6$, random couplings $J_{s,s'}\in [1/2, 3/2]$ and random fields $h_s \in [0,0.3]$.
    In b) we show the average star expectation value for the toric code with $L_x=L_y=6$, $h_{x,e}, h_{z,e}\in [0.8h,1.2h]$ and time step $\delta t = 0.05$.}
    \label{fig:big}
\end{figure}
In Fig.~\ref{fig:clock_ed} we benchmark both methods against ED in the clock model on a $4\times4$ lattice.
Here we use $\chi\in\{8,16,32,64\}$, where $\chi=64$ is close to the largest bond dimension that fits into the  GPU  memory.
Deep in the DTC phase ($\varepsilon=0.03$), both methods are accurate at small resources.
For $\varepsilon =0$, the kick operator acts as a perfect clock shift and hence transforms a product state into another product state. 
Similarly, in the Heisenberg picture, a single Weyl string remains a single Weyl string.
Thus for $\varepsilon=0.03$ the state stays efficiently representable with a low bond dimension and the observable stays relatively sparse in the Weyl basis. SWP exhibits an error below $\sim 2 \cdot 10^{-3}$ for $K=3^{12}$ (runtime of $\sim \qty{1}{\second}$) and BP-PEPS an error below $10^{-5}$ over 40 Floquet periods (Fig.~\ref{fig:clock_ed}b).

The cheapest BP-PEPS run ($\chi=8$) takes $\sim \qty{2}{\minute}$ while SWP with $K = 3^{17}$ takes $\sim \qty{5}{\minute}$.
In contrast to the precise computation of both methods in the DTC phase, in the melted phase ($\varepsilon=0.35$) the error produced by BP-PEPS saturates at an error of $\sim 10^{-1}$ (Fig.~\ref{fig:clock_ed}c)). 
Increasing the bond dimension from $\chi = 16$ up to $\chi = 64$ increases the accuracy of the BP-PEPS computations only slightly.
This is in agreement with the fact that entanglement builds up rapidly in the melted phase, so that increasing the bond dimension does not yield a big improvement. We also implemented loop corrections~\cite{vqks-cr6x} for BP-PEPS but found no significant improvement in the accuracy.
While the entanglement build-up is a limitation for the BP-PEPS simulations, SWP remains more accurate with a maximum error of $\sim10^{-2}$ at around 10 Floquet periods, while the error decreases to around $\sim10^{-3}$ at 40 Floquet periods. 
This reflects the central premises that although the state becomes highly entangled, the observable, evolved in the Heisenberg picture, can be well approximated by a sparse linear combination of Weyl strings.

For the $3 \times 3$ toric code simulations, we use ${\chi \in \{ 16, 32, 48\}}$.
The simulations with $\chi=8$ become numerically unstable.
For $h=0.1$, SWP is approximately three orders of magnitude more accurate than BP-PEPS, even with a cutoff of $K=3^{12}$, which corresponds to a runtime of $\sim \qty{5}{\second}$ (Fig.~\ref{fig:toric_ed}b).
For $h=1$ (Fig.~\ref{fig:toric_ed}c) both methods show significant errors but SWP produces results approximately one order of magnitude more accurate at $K=3^{18}$ with a runtime of $\sim \qty{50}{\minute}$ compared to a runtime of $\sim \qty{37}{\hour}$ with BP-PEPS with $\chi=48$.

The system sizes of the $8\times8$ clock model and $6 \times 6$ toric code in Fig.~\ref{fig:big} are out of reach for ED, since the Hilbert space dimensions are $3^{64}$ and $3^{72}$, respectively. 
Thus we do not have exact results as a reference to benchmark against. 
In the absence of exact reference data, we assess the accuracy by extrapolating the qualitative trends observed for smaller systems and by comparing with physical expectations.
First, in the melted phase ($\varepsilon=0.35$) of the clock model (Fig.~\ref{fig:big}a) the BP-PEPS calculations, again, saturate at $Z_3\approx 0.1$, while the SWP achieves a value of $Z_3 \approx 0.03$ consistent with the ED and SWP in the $4 \times 4$ case.
In the toric code simulations (Fig.~\ref{fig:big}b), the results from SWP remain close to $ S\approx 1$ for $h=0.1$, while the BP-PEPS computations yield results drifting away from this value to around $S \approx 0.4$. 
We expect the SWP calculations to be more precise, for two reasons. 
First, the system in reach for ED (Fig.~\ref{fig:toric_ed}) showed the same behavior with a value at around $S\approx 1$ for the weak quench.
Second, the quench field of $h=0.1$ is weak in comparison to the energy required to create a charge-anticharge-pair with an excitation gap of $\Delta = 3$, so we expect the deviations from 1 to be minor \cite{Schulz_2012}.
For the strong quench $h=1$, the SWP results are not fully converged at intermediate times ($t\sim2$), but approach the same value for later times and saturate at $S(t) \approx 0.3$.
The largest SWP simulation and BP-PEPS simulation in Fig.~\ref{fig:big} (Toric $6\times6$ with $h=1.0$) take about $\qty{3.3}{\hour}$ ($K=3^{18}$) and $\qty{6.1}{\hour}$ ($\chi=16$), respectively.

%% file: content/discussion_outlook.tex
\textit{Discussion and outlook}--We have generalized sparse Pauli propagation to qudits by propagating observables in the Weyl basis, based on a closed-form update rule under which a Weyl string branches into at most $d$ Weyl strings. 
Benchmarks on the $\mathbb{Z}_3$ clock model and the $\mathbb{Z}_3$ toric code show that sparse Weyl propagation can efficiently and accurately capture the time evolution of observables, in regimes where BP-PEPS fails, e.g., in the melted phase of the two-dimensional $\mathbb{Z}_3$ clock model and after quenches of the toric code, while both methods perform well in the DTC phase.
Qudit processors have recently realized closely related settings, including a $\mathbb{Z}_3$ toric-code ground state~\cite{iqbal2025}, period-tripling time crystals~\cite{goss2026,camacho2024} and two-dimensional lattice gauge theories~\cite{meth2025}.
SWP provides a classical reference for such experiments at system sizes beyond the reach of ED.
Several directions are interesting to explore for SWP.
The update rule holds for arbitrary local dimensions $d$, which remain to be explored numerically for $d>3$, especially in comparison to tensor network methods designed specifically for large local Hilbert spaces \cite{SciPostPhys.10.3.058}.
Extensions to noisy circuits~\cite{fb28-wlv2, PhysRevLett.133.120603} allows sparse Weyl propagation to realistically describe decoherence in qudit experiments. Finally, the recently proposed combination of SPP with quantum computation\cite{Fuller2026} can naturally be extended to qudit quantum processors with SWP.

%% file: content/appendix.tex
\section{Derivation of the update rule}
\label{app:update}
To derive the closed form of the update rule and the accompanying coefficients, we first derive the coefficients $f_m$ in the expansion of the unitary  $U = \exp(-i\theta(\exp(i\vartheta)W+ \exp(-i\vartheta)W^\dagger)) = \sum_{p=0}^{d-1} f_p W^p$ with $\vartheta,\theta \in \mathbb{R}$ as a polynomial in the generator $W$.
We use $W \equiv W_{(\mathbf{x}, \mathbf{z})}$ as long as it is unambiguous for ease of notation.
Since $W$ is unitary, it is diagonalizable, and $U$, being a function of $W$, is diagonal in the same basis.
Then using a spectral decomposition yields
\begin{equation}
    U = \sum_{q=0}^{d-1} e^{-i\theta(\exp(i\vartheta)\mu_q +\exp(-i\vartheta)\mu_q^*)} P_q, \label{eqn:U_spectral}
\end{equation}
where $\mu_q$ is the eigenvalue of $W$ in the eigenspace the projector $P_q$ projects onto.
The fact that there may only be $d$ distinct eigenvalues will follow shortly.
To compute the eigenvalues $\mu_q$ we observe that $W^d\ket{q} =  \mu_q^d \ket{q}$, where $\ket{q}$ is an eigenstate of W to the eigenvalue $\mu_q$.
To compute $W^d$, one needs to note that for a general Weyl string $W_{(\mathbf{x}, \mathbf{z})}^2 = \omega^{\mathbf{x}\cdot \mathbf{z }} W_{2\mathbf{x}, 2 \mathbf{z}}$ holds. 
By induction, one obtains $W^l = \omega^{\sum_{j=0}^{l-1} j \mathbf{x}\cdot \mathbf{z}} W_{l\mathbf{x}, l \mathbf{z}}$,
where $\sum_{j=0}^{l-1} j = l(l-1)/2$.
For $l = d$, we get $W_{(\mathbf{x}, \mathbf{z})}^d = \omega^{d(d-1)/2\,\mathbf{x}\cdot \mathbf{z}}\openone $, since a single clock or shift operator to the power of $d$ yields unity.
For odd $d$, the exponent in the phase factor yields an integer multiple of $d$, since $d-1$ is even and an even number divided by two yields an integer, as does $\mathbf{x}\cdot \mathbf{z}$.
That means that the phase yields 1 for odd $d$.
For even local dimensions $d$, it is convenient to write the phase
as $(\omega^{d/2\, \mathbf{x}\cdot \mathbf{z}})^{d-1}$.
Note that  $\omega^{d/2\, \mathbf{x}\cdot \mathbf{z}} = \exp(i\pi\mathbf{x} \cdot \mathbf{z})$.
For even $\mathbf{x} \cdot \mathbf{z}$ this yields 1 while for odd $\mathbf{x} \cdot \mathbf{z}$ it yields -1.
Since $d-1$ is always odd for even $d$, we get  
\begin{equation}
  W_{(\mathbf{x}, \mathbf{z})}^d =
  \begin{cases}
    \openone,  & \text{if } d \text{ odd},\\
    (-1)^{\mathbf{x}\cdot \mathbf{z}}\openone, & \text{if } d \text{ even}.
  \end{cases}
\end{equation}
Thus the eigenvalue equation becomes 
$W_{(\mathbf{x}, \mathbf{z})}^d\ket{q} =  \mu_q^d \ket{q} = (-1)^{(d-1)\mathbf{x}\cdot\mathbf{z}} \ket{q}$ and we arrive at $\mu_q^d - (-1)^{(d-1)\mathbf{x}\cdot\mathbf{z}} = 0$.
Since this equation is a polynomial of $d$-th order in $\mu_q$, it has $d$ solutions, which all lie on the unit circle
$\mu_q = \exp(i2\pi(q+\delta)/d) = \omega^{q+\delta}$, where we introduced $\delta  = 1/2$ if $d$ is even and $\mathbf{x}\cdot\mathbf{z}$ is odd, $\delta = 0$ otherwise.
The projectors $P_q$ are constructed as a Fourier transform of $W$, 
$P_q =1/d \sum_{p=0}^{d-1} \omega^{-p(q+\delta)} W^p$, which in fact projects a component $\ket{q}$ out of an arbitrary state $\ket{\phi} = \sum_{k=0}^{d-1} \alpha_k \ket{k}$, where $\ket{k}$ belongs to the eigenspace corresponding to the eigenvalue $\mu_k$ of $W$,
\begin{equation}
    P_q \ket{\phi} = \frac{1}{d} \sum_{k=0}^{d-1} \alpha_k \sum_{p=0}^{d-1} \omega^{-p(q+\delta)} W^p \ket{k} = \frac{1}{d}\sum_{k=0}^{d-1} \alpha_k \sum_{p=0}^{d-1} \omega^{-p(q+\delta)} \omega^{p(k+\delta)} \ket{k} = \sum_{k=0}^{d-1} \alpha_k \delta_{k,q} \ket{k} = \alpha_q \ket{q}.
\end{equation}
Noting that $\exp(i\vartheta)\mu_q + \exp(-i\vartheta)\mu_q^* = 2\cos(2\pi (q+\delta)/d +\vartheta)$ and inserting into Eq.~\eqref{eqn:U_spectral} yields 
\begin{equation}
    U = \sum_{p=0}^{d-1} f_p W^p, \quad f_p = \frac{1}{d}\sum_{q=0}^{d-1}  e^{-i2\theta \cos(2\pi (q+\delta)/d+\vartheta)} \omega^{-p(q+\delta)}. \label{eqn:coeffs}
\end{equation}
The conjugation of an arbitrary Weyl string $V_{(\mathbf{x}', \mathbf{z}')}$ by a unitary gate ${U_{(\mathbf{x}, \mathbf{z})} = \exp(-i\theta(\exp(i\vartheta)W_{(\mathbf{x}, \mathbf{z})}+\exp(-i\vartheta)W_{(\mathbf{x}, \mathbf{z})}^\dagger))}$ yields 
\begin{align}
    U_{(\mathbf{x}, \mathbf{z})}^\dagger  V_{(\mathbf{x}', \mathbf{z}')} U_{(\mathbf{x}, \mathbf{z})} &= \sum_{q=0}^{d-1} \sum_{p=0}^{d-1} f_p^* f_q W_{(\mathbf{x}, \mathbf{z})}^{-p} V_{(\mathbf{x}', \mathbf{z}')} W_{(\mathbf{x}, \mathbf{z})}^q \\
    &= \sum_{q=0}^{d-1} \sum_{p=0}^{d-1} f_p^* f_q \omega^{-p(\mathbf{x}'\mathbf{z} - \mathbf{x}\mathbf{z}')} V_{(\mathbf{x}', \mathbf{z}')} W_{(\mathbf{x}, \mathbf{z})}^{q-p}  \\
    &= \sum_{m=0}^{d-1} \sum_{p=0}^{d-1} f_p^* f_{(p+m)\bmod  d} \, \omega^{-p(\mathbf{x}'\mathbf{z} - \mathbf{x}\mathbf{z}')}
    \gamma_{(\mathbf{x},\mathbf{z})}^{ \lfloor (p+m)/d\rfloor}
    V_{(\mathbf{x}', \mathbf{z}')} W_{(\mathbf{x}, \mathbf{z})}^{m}, 
\end{align}
where $\gamma_{(\mathbf{x},\mathbf{z})} = 1$ for odd $d$ and 
$\gamma_{(\mathbf{x},\mathbf{z})} = (-1)^{\mathbf{x}\cdot\mathbf{z}}$ for even $d$.
The notation  $\lfloor y \rfloor$ means that $y$ is rounded down to the nearest integer.
In the last equation, we used $q = (p+m)\bmod d, \; m=0,\ldots d-1$.  

\section{Simulation details}
\label{app:simulation}
 
Exact reference data are obtained with a matrix-free state vector simulator on a GPU with double precision, which applies the gates on the full $3^{N_e}$-dimensional state-vector.

For the tensor network simulations on a single GPU with single precision, we use the Julia package \texttt{TensorNetworkQuantumSimulator.jl}~\cite{rudolph2024}. 
Here, we represent the state as a PEPS with maximum bond dimension $\chi$. As in the simple-update scheme, gates are applied locally and the resulting bonds are truncated down to the maximum bond dimension $\chi$ in an environment obtained from BP. 
Expectation values are obtained with the same BP environment. 
Here, the errors arise from the truncation to the maximum bond dimension $\chi$ and the BP approximation.
Here, the errors arise from the truncation to the maximum bond dimension $\chi$ and the BP approximation.
Here, the errors arise from the truncation to the maximum bond dimension $\chi$ and the BP approximation.

For the SWP, a Weyl string is stored as two bit strings encoding the exponents $\mathbf{x}$ and $\mathbf{z}$, with two bits per qutrit for each exponent $x_i, z_i \in \{ 0,1,2 \}$.
The SWP simulations are run on a single GPU with double precision.

Simulations were performed on a workstation with two AMD EPYC 9454 processors
(48 cores each, 2.75\,GHz base clock), 1.5\,TB of RAM, and two NVIDIA RTX 6000
Ada Generation GPUs (48\,GB of memory each). 
The 48\,GB memory of the GPU acts as a budget.
If a method is not converged, we increase either the maximum bond dimension for BP-PEPS simulations or the truncation parameter $K$ for the SWP simulations until we run out of GPU memory.